# A Comparison of Indonesia's E-Commerce Sentiment Analysis for Marketing Intelligence Effort
## (case study of *Bukalapak, Tokopedia* and *Elevenia*)

Andry Alamsyah [1], Fatma Saviera [2]

School of Economics and Business, Telkom University
Bandung, Indonesia
[1]andrya@telkomuniversity.ac.id, [2]fatmasaviera@gmail.com

**Abstract**

*The rapid growth of the e-commerce market in Indonesia, making various e-commerce companies appear and there has been high competition among them. Marketing intelligence is an important activity to measure competitive position. One element of marketing intelligence is to assess customer satisfaction. Many Indonesian customers express their sense of satisfaction or dissatisfaction towards the company through social media. Hence, using social media data provides a new practical way to measure marketing intelligence effort.*

*This research performs sentiment analysis using the naive bayes classifier classification method with TF-IDF weighting. We compare the sentiments towards of top-3 e-commerce sites visited companies, are Bukalapak, Tokopedia, and Elevenia. We use Twitter data for sentiment analysis because it's faster, cheaper, and easier from both the customer and the researcher side. The purpose of this research is to find out how to process the huge customer sentiment Twitter to become useful information for the e-commerce company, and which of those top-3 e-commerce companies has the highest level of customer satisfaction. The experiment results show the method can be used to classify customer sentiments in social media Twitter automatically and Elevenia is the highest e-commerce with customer satisfaction.*

## 1. Introduction

The rapid growth of the internet in Indonesia makes lifestyle changes, one of them is the increasingly popular online shopping activity. Indonesia is one of the countries with the largest e-commerce market growth in the Asia Pacific [1]. There is much e-commerce in Indonesia, but *Bukalapak, Tokopedia,* and *Elevenia* are among the top-3 popular e-commerce sites visited, their business model is Customer to Customer, as known as C2C, with an escrow payment system [2]. The effort to stay competitive among similar e-commerce sites is measured by marketing intelligence activity. Marketing intelligence represents a continuous process of understanding, analyzing, and assessing a firm's internal and external environments associated with customers, competitors, markets and then using the acquired information and knowledge to support the firm's marketing-related decisions. Marketing intelligence provides a road map of current and future trends in customers' preferences and needs, new market and segmentation opportunities, and major shifts in marketing and distribution to improve the firm's marketing planning, implementation, and control [3].

One important aspect of e-commerce business is customer satisfaction, while the buyer does not see the goods directly, the experience of other customers who already use the product will become an input during the consideration process before making a purchase. These customers interact through web technologies known as social networks, social media, or web 2.0 [4] which allows users to interact actively with each other as in everyday social life.

*Twitter* is often used as a place to express facts, opinions, and allow internet users to contribute to spreading information, *Twitter* has provided a new way to disseminate information in real-time and effectively. In this research, we use *Twitter* as the data source because it's faster, cheaper, and easier compared to other social media. For the customer, they can show what they feel about the product or service as soon as possible, for free, anytime and anywhere, while for researchers, *Twitte*r provides convenience because the data can be obtained for free, allows to obtain large amounts of data and more targeted, because the data is taken from customers who already have experience using the service.

The purpose of this research is to know how to process customer sentiment on social media *Twitter* to be useful information for the company and to know the comparison of customer satisfaction using sentiment analysis to e-commerce *Tokopedia, Bukalapak, and Elevenia* on *Twitter* with Bahasa Indonesia.

This research classifies customer sentiment into positive and negative based on tweets written by the customer to see the level of customer satisfaction at each company with the service has been received then compare between the three to know the position of each company. There are many methods of classifying text, but the *naive bayes* method used refers to some previous research, it is simple but has high accuracy and performance in text classification [5].

Several studies combine the naive bayes classifier method with *TF-IDF* weighting to obtain good results and performance. Research about social opinion polarization [6] using the *naive bayes classifier* method combined with *TF-IDF* weighting with pruning on attributes with occurrences of less than 0.83% and over 90%, yielding 94 word attributes used as models learning and accuracy of 97.42%. Research to compare *naive bayes classifier* and other methods, which is *support vector machine* is done [7], where *naive bayes classifier* yield accuracy value up to 97,86% and *support vector machine* method yield accuracy value up to 94,17%. From this research, we know *naive bayes classifier* has better performance than the *support vector machine* method.

This research also refers to previous paper [8], where the study uses the theory of sentiment analysis to determine and compare the positive and negative sentiments of popular smartphone products in Indonesia, using *sentiment analysis* based on *appraisal theory*, the results can be used as knowledge and consideration for business especially *marketing intelligence* field.

## 2. Research Method

The basic theory of this research is the *text mining* concept as it relates to searching patterns in the text by analyzing the text to obtain useful information for a specific purpose [9]. The research approach uses *sentiment analysis* to analyze people's opinions, sentiments, evaluations, appraisals, attitudes, and emotions on e-commerce as a product or service [10]. Text classification method in this research using *naive bayes classifier* with *TF-IDF* weighting, validate and evaluate on text classification using *K-fold cross-validation* and *confusion matrix*.

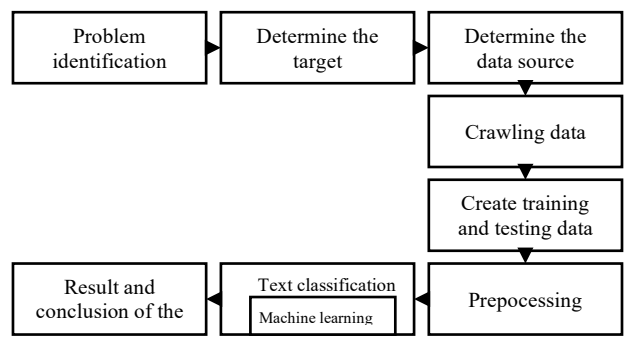

*Figure 1 The Research workflow*

The following list is the analyzing data process:

*a). Create Training and Testing Data*

Data is taken from Twitter for one week duration period from *10 to 16 October 2016*. We crawl data using *Twitter Application Programming Interface* (API) according to relevant keywords to the research. The keywords are *"Bukalapak", "Tokopedia", "Elevenia", "bukalapak_care", "tokopediacare",* and *"eleveniacare"*. Then the data is separated into training data and testing data. Training data is divided into positive and negative categories manually. From the data collection process, 69598 tweets were collected. Once the data is collected data processing and data cleaning in pre-processing step to remove irrelevant tweets like advertising, spam and others. We have as many as 2974 tweets, with 1243 tweets about *Bukalapak*, 1178 tweets about *Tokopedia,* and 553 tweets about *Elevenia*. The data is used both for training data and testing data. Of the total 2,974 tweets of data, 700 tweets were used as training data taken from each research object. The training data in this study consisted of 350 text data labeled negative and 350 labeled positive.

The following table 1 shows several example training data texts used in model construction in the study

*Table 1 Example of text data used in training data*

| Sentences | Sentiment |
|---|---|
| tidak ada tanggung jawab jawab | Negatif |
| layanan baik | Positif |
| penjual rugi tombol bantuan tidak ada | Negatif |
| promo pulsa mantap | Positif |
| cara ribet | Negatif |
| iklan iklan lucu | Positif |

*b). Cleaning and Preprocessing Data*

In the data preprocessing stage content of mined tweets will go through some process of cleaning data from noise as preparation for the basic classification process. Some preprocessing steps were performed to ensure that our algorithm works well with data. In this research, we start with *case-folding*, eliminating repetitive tweets or having duplicates, removing *URL* links and *Twitter* account names, filtering, and removing unnecessary punctuation. Also, also *tokenize, stemming, and stopword*. The whole process is shown in figure 2.

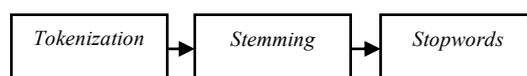

*Figure 2 Prepocessing workflow*

- *Tokenize* is a process that aims to divide the input data format is still a long text into small units called tokens. Token in the context of a document can be a word, number, or punctuation.
- *Stemming* is a technique used to find the basic word. This is based on the fact that words with the same basic form will describe the same or relatively close meaning.
- *Stopword*, in a document there are many meaningless words, such as hyphens, forward words, and others. These types of words are termed *stopwords*. In order, not to confuse the process of document processing, *stopword* should be discarded.

*c). TF-IDF Weighting*

The *TF-IDF* is a method of weighting which is integrated between the *Term Frequency* and the *Inverse Document Frequency*. *TF-IDF* will give higher weight to words that have a low frequency of occurrence in some documents, and simultaneously, has a high frequency of occurrence in a document. This ensures that words with high *TF-IDF* values can be used as representative examples of the documents in which they originate, and also words like *"the"* that often appear in documents, will be given a low weight.

The *TF-IDF* weighting process in this study uses attributes that range from 0.99% to 90% to reduce complexity and increase. Attributes are words with high *TF-IDF* values and are used as interpenetration-differentiating features. In this process, the training data model produces 86 word attributes used for the learning process in the test data

*d). Naive Bayes Classifier*

*Naïve Bayes Classifier* is a classification method that can be used to predict probabilities [11]. The *Naïve Bayes Classifier* method takes two stages in the text classification process, i.e. the training stage and the classification phase. In the training phase, the process of analyzing the sample of documents in the form of vocabulary selection, which is a word that may appear in the collection of sample documents that as far as possible can be a document representation. Next is the determination of the prior probability for each category based on the sample document. At the classification stage, the category value of a document is determined based on the term that appears in the classified document.

*e). Validation and Evaluation*

To validate and evaluate the *naïve bayes algorithm*, several tests were performed using *K-fold cross-validation* and *confusion matrix*.

In *K-fold cross-validation*, the initial data is divided randomly into a subset or fold, i.e., $D_1, D_2, D_3, ..., D_k$, each of which is estimated to be of the same size. Training and testing are performed as much as $k$ times. In iteration $i$, the data subset $D_i$ is used as the test dataset, and the remaining partition is collectively used as training data to derive a classification model to be used in the testing process. The classification model is trained and tested $k$ times. Each time trained on all but one-fold and tested on the remaining single fold. Empirical studies show a value of 10 is the optimal number of folds because the bias and variance are relatively low. Thus, the research will be validated as much as 10 times [12].

*The confusion matrix* provides an assessment of classification performance by object correctly or falsely. It also provides the decisions obtained from data training and data testing [13].

*Table 2 Confusion matrix*

|  | Classification | |
|---|---|---|
|  | **Positive** | **Negatif** |
| **Pred. Positive** | (True Positive-TP) | (False Negative-FN) |
| **Pred. Negatif** | (False Positive-FP) | (True Negative-TN) |

## 3. Result and Analysis
### 3.1 Model Evaluation

After passing the validation phase with *K-fold cross-validation* ten times, the data model yields an accuracy value of 94.57%, 94.57% recall, 94.64% Precision, and 0.891% *kappa* value.

*Accuracy* is the ratio of the exactness of the predicted results of the positive and negative classes to the total number of predicted classes. *Precision* and *Recall* are two calculations that are widely used to measure the performance of the system/system used. *Precision* is the level of accuracy between the information requested by the user and the answers provided by the system. While the *recall* is the system success rate in rediscovering information.

*Kappa* is used to measure the agreement between each pair of annotators in which the annotator is used for the classification of text classification methods [14]. The *kappa* value shows a number greater than 0.75 then the classification model can be concluded very well [15].

*Table 1 The result of the confusion matrix*

|  | Classification | | Class Precision |
|---|---|---|---|
|  | **Positive** | **Negative** |  |
| **Pred. Positive** | 338 | 26 | 92,86% |
| **Pred. Negative** | 12 | 324 | 96,43% |
| **Class Recall** | 96,57% | 92,57% |  |

### 3.2 Sentiment Analysis Results

- In *Bukalapak* data consisting of 993 text data tested on the model, it successfully classified 348 positive text data and 410 negative text data. The results show that 54.1% of customers are not satisfied with the products or services they receive.
- In *Tokopedia* data consisting of 928 text data tested on the model, it successfully classifies 350 positive text data and 408 negative text data. In Tokopedia 54.38% of customers are not satisfied with the products. or services they receive.
- Whereas in *Elevenia* data consisting of 353 text data tested on the model, it successfully classified 163 positive text data and 189 negative text data. That means 53.7% of customers are not satisfied with the products/services they receive.

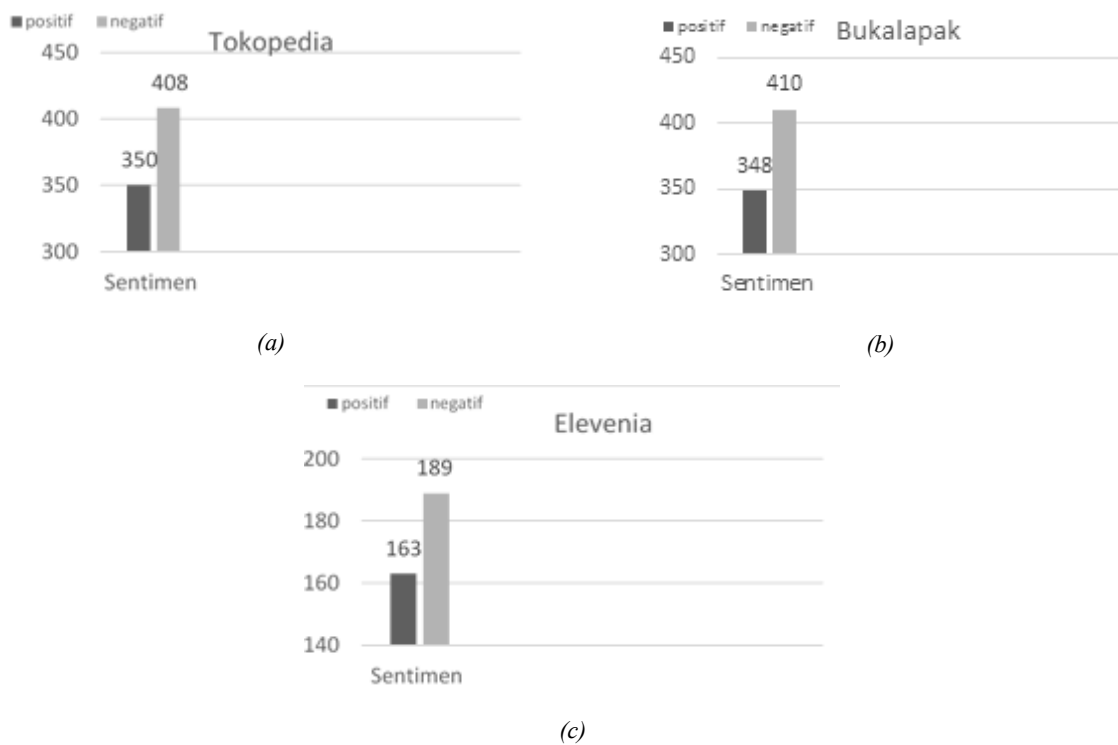

*Figure 2. (a) Bukalapak sentiment. (b) Tokopedia sentiment. (c) Elevenia sentiment*

From the results of the exposure can be seen that more customers are expressing their dissatisfaction so that when sentiments are classified the level of negative sentiment is more dominant than the level of positive sentiment in social media *Twitter*. *Elevenia* gets the highest positive sentiment level of 46.3% and negative sentiment of 53.7%. It differed slightly with *Tokopedia* which gained 46.2% positive sentiment and 53.8% negative sentiment. While *Bukalapak* obtained the lowest positive sentiment level of 45.9% and negative sentiment of 54.1%.

*Table 3 Comparison percentage sentiment among Bukalapak, Tokopedia and Elevenia on social media Twitter*

|  | Testing Data Number | Sentiment | |
| --- | --- | --- | --- |
|  |  | Positive | Negative |
| Bukalapak | 993 | 348 (45,9%) | 410 (54,1%) |
| Tokopedia | 928 | 350 (46,2%) | 408 (53,8%) |
| Elevenia | 300 | 163 (46,3%) | 189 (53,7%) |

## 4. Conclusion

Customer satisfaction is a very important aspect of the e-commerce business. Nowadays, many prospective customers search for information before making a purchase through social media, the experience of other customers who already use the product or service will be a valuable input before making a purchase.

*Twitter* gives a lot of conveniences because it can be accessed quickly, easily, and cheaply. It can be utilized by customers in conveying the facts, opinions, and sentiments they feel about the product or service company, but it also provides convenience for researchers in collecting the required data as in this research. In addition, data retrieval through *Twitter* is more appropriate because the sentiments that exist on *Twitter* is the experience of people who have used the products or services provided by the company, therefore the research can be done more effective and efficient.

Sentiment analysis using the *naive bayes classifier* method can be automatically used to classify customer sentiment accurately. It can be utilized in various ways, in this case, it uses for *marketing intelligence* in order to know how the level of customer satisfaction to the company or competitors. The result of customer satisfaction

6based on this *sentiment analysis* can be used as a reference by the company to make decisions in accomplishing strategies to increase consumer loyalty and improve the corporate image in social media Twitter.